# Adsorption and grafting on colloidal interfaces studied by scattering techniques

## - REVISED MANUSCRIPT -

Julian Oberdisse

*Laboratoire des Colloïdes, Verres et Nanomatériaux, UMR 5587 CNRS,*

*Université Montpellier II, 34095 Montpellier, France*

oberdisse@lcvn.univ-montp2.fr

Tel : (33) [0] 4 67 14 35 23

Fax : (33) [0] 4 67 14 46 37

12[th] of October 2006

**Abstract**

The adsorption of polymer and surfactant molecules onto colloidal particles or droplets in solution can be characterized non-destructively by scattering techniques. In a first part, the general framework of Dynamic Light Scattering, Small Angle Neutron and X-ray Scattering for the determination of the structure of adsorbed layers, and namely of the density profile, is presented. We then review recent studies of layers of the model polymer poly(ethylene oxide), as homopolymer or part of a block copolymer. In this field, scattering with contrast variation has been shown to be a powerful tool to obtain a detailed description of the layer structure. Adsorption of chemically more complex systems, including polyelectrolytes, polymer complexes, grafted chains and biomacromolecules are also discussed in this review, as well as surfactant adsorption.

**Figures:** 4

## 1. Introduction

Adsorption and grafting of polymer and surfactants from solution onto colloidal structures has a wide range of applications, from steric stabilisation to the design of nanostructured functional interfaces, many of which are used in industry (e.g., detergence). There are several techniques for the characterization of decorated interfaces. Scattering counts without doubt to the most powerful methods, as it allows for a precise determination of the amount and structure of the adsorbed molecules without perturbing the sample. This review focuses on structure determination of adsorbed layers on colloidal interfaces by scattering techniques, namely Dynamic Light Scattering (DLS), Small Angle Neutron and X-ray Scattering (SANS and SAXS, respectively). The important field of neutron and X-ray reflectivity is excluded, because it is covered by a review on adsorption of biomolecules on flat interfaces [1].

The colloidal domain in aqueous solutions includes particles and nanoparticles, (micro-) emulsions, and self-assembled structures like surfactant membranes, all typically in the one to one hundred nanometer range. Onto these objects, different molecules may adsorb and build layers, possibly with internal structure. We start with a review of studies concerning a model polymer, poly(ethylene oxide) (PEO), the adsorption profile normal to the surface $\Phi(z)$ of which has attracted much attention. We then extent the review to other biopolymers, polyelectrolytes, and polymer complexes, as well as to surfactant and self-assembled layers.

Adsorption isotherm measurement are the natural starting point of all studies, and whenever they are feasible, they yield independent information to be compared to the scattering results. Apart from the detailed shape of the isotherm, they give the height and position of the adsorption plateau, and thus also how much material is unadsorbed. The last point may be important for the data analysis as these molecules also contribute to the scattering.

## 2. Analysis of  scattering from decorated interfaces

Adsorbed (or grafted) layers on colloidal surfaces can be characterized quite directly by small-angle scattering. The equation describing small-angle scattering from isolated objects (for simplicity called 'particles') with adsorbed layers reads:

$$I(q) = \frac{N}{V}\left\langle \left| A_p(q) + A_l(q) \right|^2 \right\rangle + I_{inc} = \frac{N}{V}\left\langle \left| \int \Delta\rho_p(r) e^{iqr} d^3r + \int \Delta\rho_l(r) e^{iqr} d^3r \right|^2 \right\rangle + I_{inc} \qquad (1)$$

where N/V is the number density of particles, and $\Delta\rho_p$ and $\Delta\rho_l$ are the contrasts of the particle and the layer in the solvent, respectively [*2,*3,*4]. The first integral over the volume of the particles gives the scattering amplitude $A_p$ of the particles, and their intensity can be measured independently. The second integral over the volume of the layer gives the layer contribution $A_l$. The last term, $I_{inc}$, denotes the incoherent scattering background - particularly high in neutron scattering with proton-rich samples, which must be subtracted because it can dominate the layer-contribution. In eq.(1), finally, the structure factor describing particle-particle interactions is set to one, and it needs to be reintroduced for studies of concentrated colloidal suspensions [*5,*6,7,8].

Small-angle scattering with neutrons or x-rays corresponds to different contrast conditions, which makes scattering powerful and versatile, applicable to all kinds of particle-layer combinations. The great strength of SANS is that isotopic substitution gives easy access to a wide range of contrast conditions. Eq.(1) illustrates the three possible cases. If $\Delta\rho_p = 0$ ("on-contrast" or "layer contrast"), only the layer scattering is probed. Secondly, if, $\Delta\rho_l = 0$ ("particle contrast"), only the bare particle is seen, which is potentially useful to check that 'particles' (including droplets) are not modified by the adsorption process. Only in the last situation, where $\Delta\rho_p \neq 0$ and $\Delta\rho_l \neq 0$ ("off-contrast"), both terms in eq.(1) contribute. This is important for polymer layers.

Before going into modelling, one may wish to know the quantity of adsorbed matter. For small enough particles, the limiting value $I(q\rightarrow 0)$ in small angle scattering gives direct acces to this information. For homogeneous particles of volume $V_p$ in particle contrast, we obtain $I_p(q\rightarrow 0) = \Delta\rho_p^2 \Phi_p V_p$, and equivalently for layer contrast $I_l(q\rightarrow 0) = \Delta\rho_l^2 \Phi_l V_l$, where we have introduced the volume fraction of the particles $\Phi_p = N/V\ V_p$, ($\Phi_l = N/V\ V_l$ for the layer). Note that it is not important if the layer contains solvent: $V_l$ is the "dry" volume of adsorbed material, if we set $\Delta\rho_l$ to its "dry" contrast. By measuring different contrast conditions and dividing the limiting zero-angle intensities, the adsorbed quantities can be determined regardless of structure factor influence and instrument calibration, as such contributions cancel in intensity ratios [*9,*10].

The spatial extent of an adsorbed layer can be determined via the hydrodynamic radius of particles by DLS with and without the adsorbed layer, the difference being the hydrodynamic layer thickness. However, DLS does not give any information on the amount of adsorbed matter. Alternatively, with SANS or SAXS, one can determine the particle radius, with and without adsorbed layer, as well as the adsorbed amount. If the contrasts of the particle and the adsorbed material are similar, the increase in particle radius can be directly translated into the layer thickness. If the contrasts are too different, the weighting (eq.(1)) of the two contributions needs to be taken into account, e.g. with core-shell models. The simplest ones are a special case of eq.(1), with constant contrast functions $\Delta\rho(r)$. For spherically symmetric particles and adsorbed layers, the model has only four parameters (radius and contrast of particle and layer), besides the particle concentration. The particle parameters can be determined independently, whereas the other two affect I(q) differently: An increase in layer thickness, e.g., shifts the scattering function to smaller q, whereas an increase in adsorbed amount (at fixed thickness) increases the intensity. Note that the average contrast of the layer and its thickness are convenient starting points for modelling (identification of monolayers or incomplete layers), while more elaborate core-shell models use decaying shell concentrations [*5,*11,*12].

The determination of the profile $\Phi(z)$ of adsorbed polymer chains, with the z-axis normal to the surface, needs a more involved data analysis [*2,*4]. There are two routes to $\Phi(z)$. The first one is based on a measurement in "layer contrast" ($\Delta\rho_p = 0$). According to eq.(1), with $\Delta\rho_l \propto \Phi(z)$, this intensity is related to the square of the Fourier transform of $\Phi(z)$. One can then either test different profiles, or try to invert the relationship, which causes the usual problems related to data inversion (limited q-range, phase loss and limiting conditions …) [*2,*3,*4]. This route also gives a (usually small) second term in the layer-scattering, called the fluctuation term [13], which stems from deviations from the average profile. The second route is based on additional off-contrast measurements. Carrying out the square of the sum in eq.(1) gives 3 terms, $A_p^2 + A_l^2 + 2 A_p A_l$. Subtracting the bare particle and pure layer term yields the cross-term with the layer contribution $A_l$, this time without the square, which is easier to treat because the phase factor is not lost.

# 3 Review of grafting and adsorption studies by small-angle scattering and DLS

## 3.1 Structure of PEO-layers

Many studies focussing on fundamental aspects deal with the model polymer PEO, as homopolymer or part of a block copolymer [14-30]. Especially SANS has lead to a very detailed description of the structure of PEO-layers, deepening our understanding of polymer brushes, and their interaction, e.g. in colloidal stabilization. Hone et al have measured the properties of an adsorbed layer of PEO on poly(styrene)-latex (PS) [*14]. They have performed on- and off-contrast SANS experiments in order to determine the (exponential) profile $\Phi(z)$ and the weak fluctuation term, the determination of which requires a proper treatment of smearing and polydispersity. They have revisited the calculation by Auvray and de Gennes [13], and propose a one parameter description of the fluctuation term. Marshall et al have extended the adsorbed layer study of PEO on PS-latex to different molecular weights, and compare exponential, scaling-theory based and Scheutjens-Fleer self-consistent mean field theory, including the fluctuation term [**15]. Recently, the effect of electrolytes on PEO layers on silica was also investigated by DLS [16].

Concerning copolymers, Seelenmeyer and Ballauff investigate the adsorption of non-ionic $C_{18}E_{112}$ onto PS latex particles by SAXS [*17]. They used exponential and parabolic density profiles for PEO to fit the data. The adsorption of a similar non-ionic surfactant ($C_{12}E_{24}$) onto hydrophobized silica in water was studied by SANS, employing a two layer model describing the hydrophobic and hydrophilic layers [18].

On hydrocarbon and fluorocarbon emulsion droplets, layers of two triblock copolymers (pluronics F68 and F127) and a star-like molecule (Poloxamine 908) have been adsorbed by King et al [*19]. They found surprisingly similar exponentially decaying profiles in all cases, cf. Fig. 1, which also serves as illustration for two ways to determine $\Phi(z)$ in "layer contrast", inversion and fitting, as discussed in section 2. The SANS-study of Washington et al deals with small diblock copolymers adsorbed on perfluorocarbon emulsion droplets [20]. A clear temperature-dependence of $\Phi(z)$ was found, but the best-fitting profile type depends on the (low) molecular weight. Diblock copolymer layers adsorbed onto water droplets have been characterized by DLS by Omarjee et al [*21]. Frielinghaus et al determine the partial structure factors of diblock copolymers [*22,23] used for boosting of microemulsions [24].

Adsorption on carbon black has been studied by comparison of DLS and contrast-variation SANS [*25,26,27]. The adsorbed layer of both F127 and a rake-type siloxane-PPO-PEO copolymer was found to be a monolayer at low coverage, and adsorbed micelles at high coverage. On magnetic particles, Moeser et al have followed the water decrease in a PPO-PEO shell by SANS and theory as the PPO content increases [*11]. Concerning the adsorption of PEO and tri-block copolymers on non-spherical particles, Nelson and Cosgrove have performed SANS and DLS studies with anisotropic clay particles [*28,29,30]. Unusually thin layers are found for PEO, and a stronger adsorption of the pluronics.

Studies of adsorbed PEO-layers at higher colloid concentrations have been published by several groups [*5,7,8]. Zackrisson et al have studied PEO-layers grafted to PS-particles (used for studies of glassy dynamics) by SANS with contrast variation, using a stretched-chain polymer profile [*5]. In Fig. 2, they compare their form factor measurements to a model prediction, at different solvent compositions, and nice fits are obtained. Along a very different approach, Qi et al match the PEO layer but follow its influence via the interparticle structure factor [7,8].

**3.2 Structure of adsorbed and grafted layers, from polyelectrolytes to surfactants.**

**Adsorption of polyelectrolytes, biomacromolecules, and polymer complexes.**

Adsorbed layers of many different macromolecules have been characterized by scattering [31-39]. In these studies, the focus shifts from the more 'conceptual' interest in PEO-layers to specific substrate-molecule interactions. The profile of gelatin layers adsorbed on contrast matched PS-particles was shown to be well-described by an exponential by Marshall el al [31]. Addition of equally contrast matched ionic surfactant (SDS) induces layer swelling, and finally gelatin desorption. Dreiss et al have shown that the α-cyclodextrin threads on adsorbed PEO chains (pseudopolyrotaxanes), modifying their configuration [32]. Cárdenas et al have characterized DNA-coated contrast-matched PS-particles by SANS, and present evidence for layer compaction upon addition of cationic surfactant [33]. Addition of ionic surfactants has been shown to lead to the desorption of ethyl(hydroxyethyl)cellulose from PS-latex, which can be followed by DLS [34].

Scattering studies have been crucial for polyelectrolyte layers. The adsorption of small proteins (BSA) onto spherical polyelectrolyte brushes was measured by SAXS by Rosenfeldt et al [**35]. Using DLS, the thickness of adsorbed cationic copolymer on latex particles was studied by Borget et al [36]. Finally, polyelectrolyte multilayers have been characterized on contrast-matched PS using core-shell models, and by DLS on silica [**37, 38,39]. In all of these studies, unperturbed structural characterizations in the solvent were made possible by scattering.

There is a great amount of literature by synthesis groups in grafting of polymer chains onto or from colloidal surfaces. These groups often use DLS to characterize layer extensions [40,41], with convincing plots of the growing hydrodynamic thickness during polymerisation [**42], or as a function of external stimuli [*12, 43-47]. In static scattering, El Harrak et al use SANS [*48, 49, 50], and Yang et al DLS and static light scattering with a core-shell model [*12]. Concerning the structure of grafted layers, the Pedersen model must be mentioned [**51]. Shah et al use polarized and depolarised light scattering to investigate PMMA layers grafted onto Montmorillonite clay [52]. In a concentration study, Kohlbrecher et al fit contrast-variation SANS intensities of coated silica spheres in toluene with a core-shell model and an adhesive polydisperse structure factor model [*6].

**Adsorption of surfactant layers and supramolecular aggregates.**

The adsorption of ionic and non-ionic surfactants to hydrocarbon emulsion droplets is of evident industrial importance. In this case, the scattered intensity can be described by a core-shell model [*10], which was also used by Bumajdad et al to study the partitioning of C12Ej (j=3 to 8) in DDAB layers in water-in-oil emulsion droplets [53]. On colloids, the thickness of an adsorbed layer of $C_{12}E_5$ on laponite has been measured by Grillo et al by SANS using a core-shell model, and evidence for incomplete layer formation was found [54]. On silica particles, a contrast-variation study of adsorbed non-ionic surfactant has been performed, and the scattering data modelled with micelle-decorated silica [*9,55], a structure already seen by Cummins et al [56].

Pores offer the possibility to study adsorption at interfaces with curvatures comparable but opposite in sign to colloids. Porous solids are not colloidal, but adsorption inside pores can be analysed using small angle scattering. Vangeyte et al [*57] study adsorption of poly(ethylene oxide)-b-poly($\varepsilon$-caprolactone) copolymers at the silica-water interface. They succeed in explaining their SANS-data with an elaborate model for adsorbed micelles similar to bulk micelles, cf. Fig. 3, and the result in $q^2I$ representation is shown in Fig. 4. In the more complex system with added SDS the peak disappears and a core-shell model becomes more appropriate, indicating de-aggregation [58].

## 4. Conclusion

Recent advances in the study of adsorption on colloidal interfaces have been reviewed. On the one hand, DLS is routinely used to characterize layer thickness, with a noticeable sensitivity to long tails due to their influence on hydrodynamics. On the other hand, SANS and SAXS give information on mass, and mass distribution, with a higher sensitivity to the denser regions. Small-angle scattering being a 'mature' discipline, it appears that major progress has been made by using it to resolve fundamental questions, namely concerning the layer profile of model polymers. In parallel, a very vivid community of researchers makes intensive use of DLS and static scattering to characterize and follow the growth of layers of increasing complexity.

**Acknowledgements:** Critical rereading and fruitful discussions with François Boué and Grégoire Porte are gratefully acknowledged.

**References:**

[1] Lu JL, Current Opinion in Colloid and Interface Science, this volume.

[*2] Fleer GJ, Cohen Stuart MA, Scheutjens JMHM, Cosgrove T, Vincent B: *Polymers at interfaces,* 1st edition, Chapman & Hall, New York, 1993.

A not brand-new but still very useful textbook on polymers at interfaces, including techniques and theory.

[*3] *Neutrons, X-ray and Light: Scattering Methods Applied to Soft Condensed Matter*; P. Lindner, Th. Zemb; eds.; North Holland, 2002.

An excellent textbook covering many aspects of scattering used in soft condensed matter studies.

[*4] King S, Griffith P, Hone J, Cosgrove T: **SANS from adsorbed polymer layers**, *Macromol Symp* 2002, **190**:33-42

This paper reviews the scattering from adsorbed layers in general. All equations are extensively discussed, with a special focus on the determination of the polymer volume fraction profile and the polymer fluctuation term.

[*5] Zackrisson M, Stradner A, Schurtenberger P, Bergenholtz J: **Small-angle neutron scattering on a core-shell colloidal system: A contrast variation study**, *Langmuir* 2005, **21**: 10835-10845

Sterically stabilized polystyrene latex particles are studied by SANS. The outer PEO-layer is described by a pincushion model, leading to a profile proportional to $r^{-2}$. Scattering at higher concentrations is described with Percus-Yevick structure factor for polydisperse mixtures.

[*6] Kohlbrecher J, Buitenhuis J, Meier G, Lettinga MP: **Colloidal dispersions of octadecyl grafted silica spheres in toluene: A global analysis of small angle neutron scattering**

**contrast variation and concentration dependence measurements**, *J Chem Phys* 2006, **125**: 044715

A contrast-variation study of the scattering of silica spheres with a hydrophobic layer in an organic solvent is presented. The intensity is described by a core-shell model combined with a structure factor for adhesive particles, which fits all contrast situations simultaneously.

[7] Qiu D, Cosgrove T, Howe A: **Small-angle neutron scattering study of concentrated colloidal dispersions: The electrostatic/steric composite interactions between colloidal particles**, *Langmuir* 2006, **22**:6060-6067

[8] Qiu D, Dreiss CA, Cosgrove T, Howe A: **Small-angle neutron scattering study of concentrated colloidal dispersions: The interparticle interactions between sterically stabilized particles**, *Langmuir* 2005,**21**:9964-9969

[*9] Despert G, Oberdisse J: **Formation of micelle-decorated colloidal silica by adsorption of nonionic surfactant,** *Langmuir* 2003, **19**, 7604-7610

The adsorption of a non-ionic surfactant (TX-100) on colloidal silica is studied by SANS, using solvent contrast variation. The adsorbed layer is described by a model of impenetrable micelles attached to the silica bead.

[*10] Staples E, Penfold J, Tucker I: **Adsorption of mixed surfactants at the oil/water interface**, *J Phys Chem B* 2000, **104**: 606-614

The adsorption of mixtures of SDS and $C_{12}E_6$ onto hexadecane droplets in water is studied by SANS. A core-shell model is used to describe the form factor of the emulsion droplets, and coexisting micelles are modelled as interacting core-shell particles. A model-independent analysis using $I(q{\rightarrow}0)$ is used to extract information on layer composition. The results are shown to disagree with straightforward regular solution theory.

[*11] Moeser GD, Green WH, Laibinis PE, Linse P, Hatton TA: **Structure of polymer-stabilized magnetic fluids: small-angle neutrons scattering and mean-field lattice modelling**, *Langmuir* 2004, **20**:5223-5234

The layer of PAA with grafted PPO and PEO blocks bound to magnetic nanoparticles is studied by SANS. Core-shell modelling including the magnetic scattering of the core is used to determine the layer density and thickness, for different PPO/PEO ratios.

[*12] Yang C, Kizhakkedathu JN, Brooks DE, Jin F, WU C: **Laser-light scattering study of internal motions of polymer chains grafted on spherical latex particles**, *J Phys Chem B* 2004, **108**:18479-18484

Temperature-dependent Poly(NIPAM) chains grown from relatively big poly(styrene) latex ('grafting from') are studied by static and dynamic light scattering. Near the theta-temperature, the hairy-latex is described by a core-shell model with a $r^{-1}$ polymer density in the brush. The time correlation function reveals interesting dynamics at small scales, presumably due to internal motions.

[13] Auvray G, de Gennes PG, **Neutron scattering by adsorbed polymer layers,** *Europhys Lett* 1986, **2** :647-650

[*14] Hone J H E, Cosgrove T, Saphiannikova M, Obey T M, Marshall J C, Crowley T L: **Structure of physically adsorbed polymer layers measured by small-angle neutron scattering using contrast variation methods**, *Langmuir* 2002, **18**: 855-864

Combined on- and off-contrast SANS experiments in order to determine the (exponential) profile $\Phi(z)$ and the weak fluctuation term of PEO-layers on polystyrene latex. The fluctuation term is obtained by subtraction of layer intensities obtained via the two routes discussed in the text.

[**15] Marshall J C, Cosgrove T, Leermakers F, Obey T M, Dreiss C A: **Detailed modelling of the volume fraction profile of adsorbed layers using small-angle neutron scattering**, *Langmuir* 2004, **20**: 4480-4488

The structure of the adsorbed layers of PEO (10k to 634 k) on polystyrene is studied by on- and off-contrast SANS. Different theoretical profiles are reviewed and used to describe the layer scattering. This includes the weak fluctuation term, which is proportional to $q^{-4/3}$ and decays more slowly than the average layer contribution $q^{-2}$. It is therefore more important (but nonetheless very small) at high q.

[16] Flood C, Cosgrove T, Howell I, Revell P: **Effect of electrolyte on adsorbed polymer layers: poly(ethylene oxide) – silica system**, *Langmuir* 2006, **22**: 6923-6930

[*17] Seelenmeyer S, Ballauff M: **Analysis of surfactants adsorbed onto the surface of latex particles by small-angle x-ray scattering**, *Langmuir* 2000, **16**: 4094-4099

The layer structure of hydrophilic PEO attached onto latex by hydrophobic stickers is studied. The PS-latex is virtually matched by the solvent, and the intensity curves show a nice evolution the side maxima, which shift to smaller-q and raise in intensity as the layer scattering increases. Moments of the density profile are used to characterize the layer, and both an exponential and a parabolic density profile fit the data.

[18] Dale P J, Vincent B, Cosgrove T, Kijlstra J: **Small-angle neutron scattering studies of an adsorbed non-ionic surfactant ($C_{12}E_{24}$) on hydrophobised silica particles in water**, *Langmuir* 2005, **21**: 12244 - 12249

[*19] King S, Washington C, Heenan R: **Polyoxyalkylene block copolymers adsorbed in hydrocarbon and fluorocarbon oil-in-water emulsions**, *Phys Chem Chem Phys* 2005, **7**:143-149

The volume profiles of three copolymers (F68, F127, tetronic/poloxamine 908) adsorbed onto emulsion droplets are determined by SANS, using two methods, inversion and fitting, to Φ(z). Considerable similarity in the adsorbed layer structure is found for hydrocarbon and fluorocarbon emulsions.

[20] Washington C, King S M, Attwood D, Booth C, Mai S M, Yang Y W: **Polymer bristles: Adsorption of low molecular weight poly(oxyethylene)-poly(oxybutylene) diblock copolymers on a perfluorocarbon emulsion**, *Macromolecules* 2000, **33**: 1289 – 1297

[*21] Omarjee P, Hoerner P, Riess G, Cabuil V, Mondain-Monval O: **Diblock copolymers adsorbed at a water-oil interface**, *Eur Phys J E* 2001, **4**: 45-50

Adsorbed layers of polybutadiene-PEO block copolymers on water-in-oil droplets are studied by DLS and by force-distance profiles based on magnetic control via included ferrofluid particles.

[*22] Frielinghaus H, Byelov D, Allgaier J, Richter D, Jakobs B, Sottmann T, Strey R, **Efficiency boosting and optional viscosity tuning in microemulsions studied by SANS**, *Appl Phys A* 2002, **74**: 408-410

The boosting effect of PPO-PEO diblock copolymers consists in an important shift of the emulsification efficiency in bicontinuous microemulsions. In this study, the partial scattering functions between the copolymer and the surfactant film, the oil and the film, and the oil and the polymer are determined by contrast variation.

[23] Frielinghaus H, Byelov D, Allgaier J, Gompper G, Richter D, **Efficiency boosting and optional viscosity tuning in microemulsions studied by SANS**, *Physica B* 2004, **350**: 186-192

[24] Jakobs B, Sottmann T, Strey R Allgaier J, Willner L, Richter D, **Amphiphilic block copolymers as efficiency boosters for microemulsions**, *Langmuir* 1999, **15**: 6707-6711

[*25] Lin Y, Alexandridis P, **Temperature-dependent adsorption of pluronic F127 block copolymers onto carbon black particles dispersed in aqueous media**, *J Phys Chem B* 2002, **106:**10834-10844

The adsorption of PEO-PPO-PEO (F127) block copolymers onto carbon black is studied by DLS and SANS. The hydrodynamic thickness of the adsorbed layers is found to increase with polymer concentration. Neutron scattering with contrast matching is used to verify that neither carbon black nor the micelles are modified by the other component.

[26] Lin Y, Smith T W, Alexandridis P, **Adsorption of a polymeric siloxane surfactant on carbon black particles dispersed in mixtures of water with polar organic solvents**, *J Coll Int Sci* 2002, **255:**1-9

[27] Lin Y, Smith T W, Alexandridis P, **Adsorption of a rake-type siloxane surfactant onto carbon black nanoparticles dispersed in aqueous media**, *Langmuir* 2002, **18**:6147-6158

[*28] Nelson A, Cosgrove T: **A small-angle neutron scattering study of adsorbed poly(ethylene oxide) on laponite**, *Langmuir* 2004, **20**: 2298-2304

The adsorption of PEO between 20 and 276 k on clay particles is studied by NMR solvent relaxtion and by SANS. Core-shell modelling is used to obtain adsorbed mass. The layer thickness is found to be very thin and smaller on the face and the edge of the clay (16 Å on the face, 35Å on the edge).

[29] Nelson A, Cosgrove T: **Dynamic light scattering studies of poly(ethylene oxide) adsorbed on laponite: Layer conformation and its effect on particle stability**, *Langmuir* 2004, **20**: 10382-10388

[30] Nelson A, Cosgrove T: **Small-angle neutron scattering study of adsorbed pluronic tri-block copolymers on laponite**, *Langmuir* 2005, **21**: 9176-9182

[31] Marshall J C, Cosgrove T, Jack K: **Small-angle neutron scattering of gelatin/sodium dodecyl sulfate complexes at the polystyrene/water interface**, *Langmuir* 2002, **18**: 9668-9675

[32] Dreiss C A, Cosgrove T, Newby F N, Sabadini E: **Formation of a supramolecular gel between α-cyclodextrin and free and adsorbed PEO on the surface of colloidal silica: Effect of temperature, solvent, and particle size**, *Langmuir* 2004, **20**: 9124-9129

[33] Cárdenas M, Dreiss C A, Nylander T, Chan C P, Cosgrove T, Lindman B: **SANS study of the interactions among DNA, a cationic surfactant, and polystyrene latex particles**, *Langmuir* 2005, **21**: 3578-3583

[34] Lauten R A, Kjøniksen AL, Nyström B: **Adsorption and desorption of unmodified and hydrophobically modified ethyl(hydroxyethyl)cellulose on polystyrene latex particles in the presence of ionic surfactants using dynamic light scattering**, *Langmuir* 2000, **16**:4478-4484

[**35] Rosenfeldt S, Wittemann A, Ballauff M, Breininger E, Bolze J, Dingenouts N: **Interaction of proteins with spherical polyelectrolyte brushes in solution as studied by small-angle x-ray scattering,** *Phys Rev E* 2004, **70**: 061403

The adsorption of Bovine Serum Albumin (BSA) and Bovine Pancreastic Ribonuclease A (RNase A) on strong and weak polyelectrolytes grafted onto latex particles is measured by SAXS. The scattered intensity is modelled by a geometrical model of beads on the surface of the latex, cf. [*9, *57].

[36] Borget P, Lafuma F, Bonnet-Gonnet C, **Interactions of hairy latex particles with cationic copolymers,** *J Coll Int Sci* 2005, **284:**560-570

[**37] Estrela-Lopis I, Leporatti S, Moya S, Brandt A, Donath E, Möhwald H: **SANS studies of polyelectrolyte multilayers on colloidal templates** *Langmuir* 2002, **18**:7861-7866

The adsorption of a polyelectrolyte multiplayer on index-matched polystyrene latex is studied by SANS using a core-shell model. This allows to deduce the average thickness of a polyelectrolyte layer (16Å), and the amount of water inside the layers.

[38] Rusu M, Kuckling D, Möhwald H, Schönhoff M, **Adsorption of novel thermosensitive graft-copolymers: Core-shell particles prepared by polyelectrolyte multiplayer self-assembly,** *J Coll Int Sci* 2006, **298:**124-131

[39] Okubo T, Suda M, Tsuchida A: **Alternate multi-layered adsorption of macro-cations and -anions on the colloidal spheres. Influence of the deionisation of the complexation mixtures with coexistence of the ion-exchange resins**, *Coll Polym Sci* 2005, **284**:284-292

[40] Inoubli R, Dagréou S, Khoukh A, Roby F, Peyrelasse J, Billon L : **'Graft from' polymerization on colloidal silica particles: elaboration of alkoxyamine grafted surface by in situ trapping of carbon radicals**, *Polymer* 2005, **46**:2486–2496

[41] Qi DM, Bao YZ, Weng ZX, Huang ZM: **Preparation of acrylate polymer/silica nanocomposite particles with high silica encapsulation efficiency via miniemulsion polymerisation**, Polymer 2006, **47**:4622-4629

[**42] Bartholome C, Beyou E, Bourgeat-Lami E, Chaumont P, Lefebvre F, Zydowicz N: **Nitroxide-mediated polymerization of styrene initiated from the surface of silica nanoparticles. In situ generation and grafting of alkoxyamine initiators**, *Macromolecules* 2005**, 38**: 1099-1106

Poly(styrene) chains are grown from the surface of silica particles ('grafting from'), with a thorough characterization and control of the chemical reaction. This allows the measurement the hydrodynamic thickness of the layer as a function of polymer molar mass.

[43] Li D, Jones GL, Dunlap JR, Hua F, Zhao B: **Thermosensitive hairy hybrid nanoparticles synthesized by surface-initiated atom transfer radical polymerisation**, *Langmuir* 2006, **22**:3344-3351

[44] Zhang Y, Luo S, Liu S: **Fabrication of Hybrid Nanoparticles with Thermoresponsive Coronas via a Self-Assembling Approach**, *Macromolecules* 2005, **38***: 9813-9820

[45] Zhang M, Liu L, Zhao H, Yang Y, Fu G, He B: **Double-responsive polymer brushes on the surface of colloid particles**, *J Coll Int Sci* 2006, **301**:85-91

[46] Luo S, Xu J, Zhang Y, Liu S, Wu C: **Double hydrophilic block copolymer monolayer protected hybrid gold nanoparticles and their shell cross-linking,** *J Phys Chem B* 2005, **109**: 22159-22166

[47] Tan BH, Tam KC, Lam YC, Tan CB: **Microstructure and rheological properties of pH-responsive core–shell particles**, *Polymer* 2005, **46**:10066-10076

[*48] Carrot G, El Harrak A, Oberdisse J, Jestin J, Boué F; **Detailed structural observation of nanografting chemistry with neutron scattering**, *Soft Matter,* in press

The initiation and polymerization of butyl methacrylate from colloidal silica is followed by SANS. By using different solvent contrasts, the build-up of the layer and the dispersion of the colloid are followed independently.

[49] El Harrak A, Carrot G, Oberdisse J, Eychenne-Baron Ch, Boué F: **Surface-Atom Transfer Radical Polymerization from Silica Nanoparticles with Controlled Colloidal Stability**, *Macromolecules* 2004, **37**:6376-6384

[50] El Harrak A, Carrot G, Oberdisse J, Jestin J, Boué F: **Atom Transfer Radical Polymerization from Silica Nanoparticles using the "Grafting From" Method and Structural Study via Small-Angle Neutron Scattering**, *Polymer* 2005, **46**:1095-1104

[**51] Pedersen JS, Carsten Svaneborg C, Almdal K, Hamley IW, Young RN: **A small-angle neutron and x-ray contrast variation scattering study of the structure of block copolymer micelles: Corona shape and excluded volume interactions**, *Macromolecules* 2003, **36**:416-433

SANS and SAXS data of poly(styrene)-poly(isoprene) block copolymers in decane are reported in this article. The PS-core is covered by a PI layer, which is made visible by using deuterated material. Detailed modelling based either on non-interacting gaussian chains or on scattering functions based on Monte-Carlo simulations for self-avoiding chains is used to extract micellar parameters. Further work on this subject by the same group is cited in the references.

[52] Shah D, Fytas G, Vlassopoulos D, Di J, Sogah D, Giannelis EP: **Structure and dynamics of polymer-grafted clay suspensions**, *Langmuir* 2005, **21**:16-25

[53] Bumajdad A, Eastoe J, Nave S, Steytler D C, Heenan R K, Grillo I: **Compositions of mixed surfactant layers in microemulsions determined by small-angle neutron scattering**, *Langmuir* 2003, **19**: 2560-2567

[54] Grillo I, Levitz P, Zemb T: **SANS structural determination of a nonionic surfactant layer adsorbed on clay particles**, *Eur Phys J B* 1999, **10**, 29-34

[55] Oberdisse J: **Small Angle Neutron Scattering and Model predictions for Micelle-decorated Silica Beads**, *Physical Chemistry Chemical Physics* 2004, **6**(7): 1557-1561

[56] Cummins PG, Staples E, Penfold J: **Temperature-dependence of the adsorption of hexaethylene glycol monododecyl ether on silica sols**, *J Phys Chem* 1991, **95**:5902-5905

[*57] Vangeyte P, Leyh B, Rojas O J, Claesson P M, Heinrich M, Auvray L, Willet N, Jérôme R: **Adsorption of poly(ethylene oxide)-b-poly(ε-caprolactone) copolymers at the silica-water interface**. *Langmuir* 2005, **21**: 2930-2940

The behaviour of poly(ethylene oxide)-b-poly(ε-caprolactone) copolymers at the silica-water interface inside porous silica is studied.  A simple layered structure in the silica pores is ruled out, as none of the profiles explains a scattering peak at intermediate q, which can be described by a quite involved scattering model for adsorbed micelles. The model reproduces the peak by inter-micellar interactions within the layer, and is consistent with the adsorption isotherm (cooperative adsorption and plateau value).

[58] Vangeyte P, Leyh B, De Clercq C, Auvray L, Misselyn-Bauduin A M, Jérôme R: **Concomitant adsorption of poly(ethylene oxide)-b-poly(ε-caprolactone) copolymers and sodium dodecyl sulfate at the silica-water interface**. *Langmuir* 2005, **21**: 7710-7716

**Figure Captions:**

Figure 1:    Comparison of the calculated volume fraction profiles for three copolymers. The SANS data from which these are derived is from fluorocarbon emulsions at 25°C in the absence of electrolyte. The points were generated by mathematical inversion, and the lines by profile fitting assuming exponential profiles. (♦) and (——) Poloxamer 188, (■) and (– – – –) Poloxamer 407, (○) and (–· – · – · –) Poloxamine 908. Reprinted from ref. [19], reproduced by permission of the PCCP Owner Societies.
*Figure 1 is Fig. 1 of ref. [19].*

Figure 2:    Form factors measured for deuterated latex with grafted PEO-layers in 0.4M $Na_2CO_3$ at three different contrasts corresponding to 100:0, 91:9, and 85:15 (w/w) $D_2O/H_2O$. Lines are "simultaneous" fits (cf. [*5]) in which only the solvent scattering length density varies. Shown in the inset are accompanying scattering length density profiles. Reprinted with permission from ref. [*5], copyright 2005, American Chemical Society.
*Figure 2 is Fig. 7 of ref. [*5].*

Figure 3:    Micellar model used by Vangeyte et al [*57] for the analysis of the SANS intensity. Spherically symmetric micelles are assumed, consisting of a dense hydrophobic PCL core (radius *R*) surrounded by rodlike PEO chains of an effective length *L*. Similar models have been proposed in refs. [*9,**35]. Reprinted with permission from ref. [*57], copyright 2005, American Chemical Society.
*Figure 3a is Fig. 9 of ref. [*57].*

Figure 4:    Fit of the micellar form factor for the core-rigid rods model to the SANS intensity for the $PEO_{114}$-*b*-$PCL_{19}$ copolymer at surface saturation in porous silica, see ref. [*57] for details. The representation of $q^2I(q)$ vs q enhances the layer scattering. Reprinted with permission from ref. [*57], copyright 2005, American Chemical Society.
*Figure 3b is the upper graph of Fig.10 in ref. [*57].*

**Figures :** **I have copy-pasted them from the original pdf, is this ok ?**

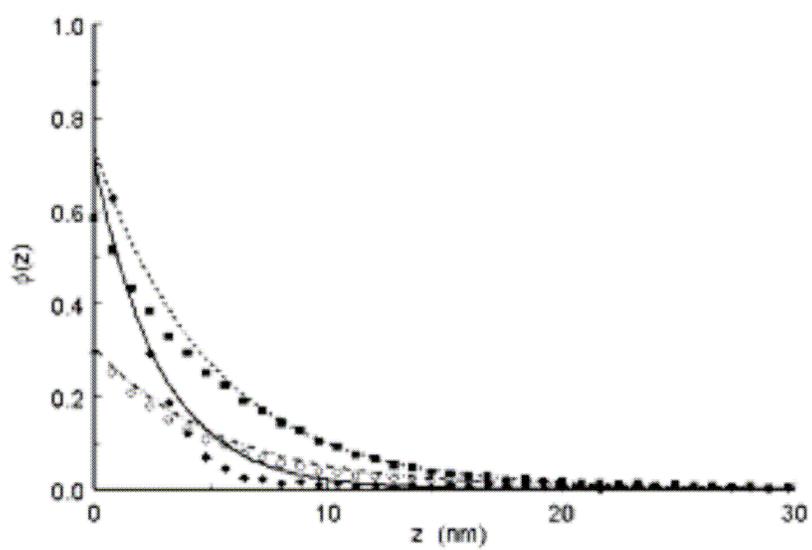

Figure 1 (Oberdisse)

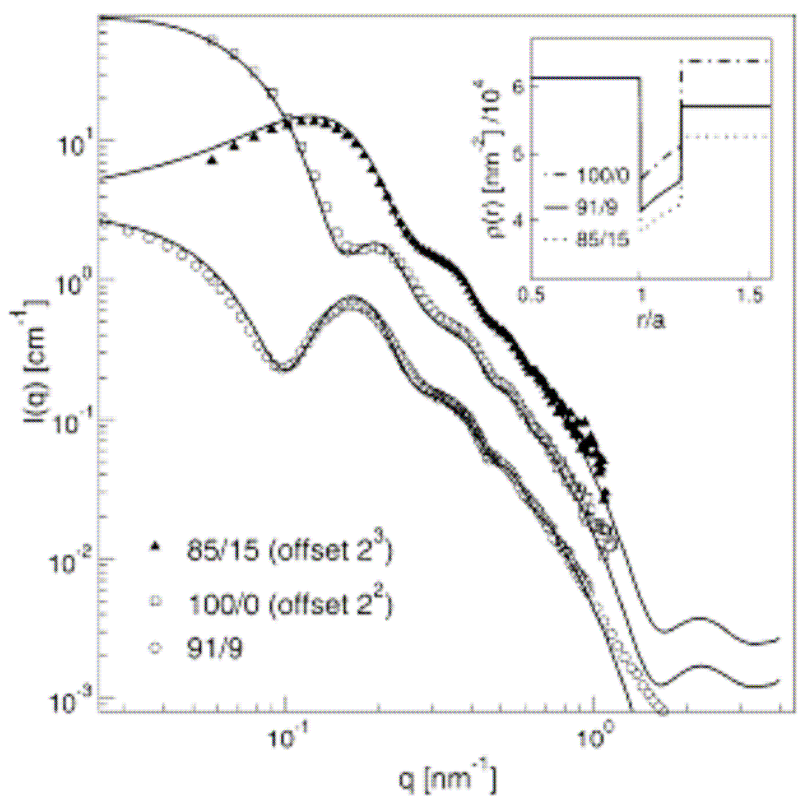

Figure 2 (Oberdisse)

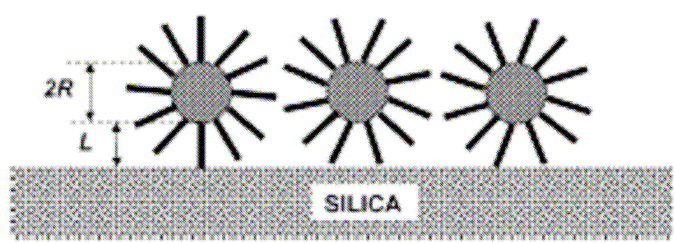

Figure 3 (Oberdisse)

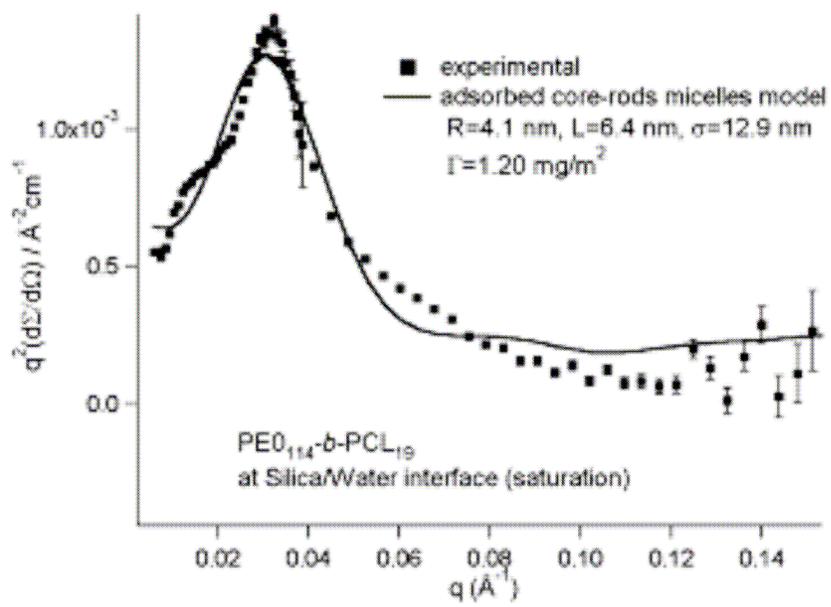

Figure 4 (Oberdisse)